\documentclass[prl,twocolumn,superscriptaddress,showkeys]{revtex4}
\usepackage{amssymb,epsf,epsfig,amsmath,color}

\newcommand{\be}{\begin{equation}}
\newcommand{\ee}{\end{equation}}

\newcommand{\lsim}{
\mathrel{\hbox{\rlap{\hbox{\lower4pt\hbox{$\sim$}}}\hbox{$<$}}}}
\newcommand{\gsim}{
\mathrel{\hbox{\rlap{\hbox{\lower4pt\hbox{$\sim$}}}\hbox{$>$}}}}

\newcommand{\beq}{\begin{equation}}
\newcommand{\eeq}{\end{equation}}

\begin{document}
\begin{titlepage}
\vspace*{1.7truecm}
\begin{flushright}
CERN-PH-TH/2008-133
\end{flushright}

\vspace{1.6truecm}

\begin{center}
\boldmath
{\Large{\bf Benchmarks for the New-Physics Search through\\
\vspace*{0.3truecm}
CP Violation in $B^0\to\pi^0 K_{\rm S}$}}
\unboldmath
\end{center}

\vspace{1.2truecm}

\begin{center}
{\bf Robert Fleischer,${}^a$
Sebastian J\"ager,${}^a$ Dan Pirjol${}^b$ and Jure Zupan${}^{a,c,d}$}

\vspace{0.5truecm}

${}^a$ {\sl Theory Division, Department of Physics, CERN, CH-1211 Geneva 23,
Switzerland}

\vspace{0.2truecm}

${}^b$ {\sl National Institute for Physics and Nuclear Engineering, 
Department of Particle Physics,\\ 077125 Bucharest, Romania}

${}^c${\sl J.~Stefan Institute, Jamova 39, 1000 Ljubljana, Slovenia}

${}^d${\sl Faculty of mathematics and physics, University of Ljubljana, Jadranska 19, 1000 Ljubljana, Slovenia}

\end{center}

\vspace*{1.7cm}

\begin{center}
\large{\bf Abstract}

\vspace*{0.6truecm}

\begin{tabular}{p{14.5truecm}}
{\small 
Using isospin relations, we predict the Standard-Model correlation between 
$S_{\pi^0 K_{\rm S}}\equiv (\sin 2\beta)_{\pi^0 K_{\rm S}}$ and $A_{\pi^0 K_{\rm S}}$, the mixing-induced and direct CP asymmetries of $B^0\to \pi^0 K_{\rm S}$. The 
calculation uses flavour $SU(3)$ only to fix the isospin-3/2 amplitude through the 
$B^\pm\to\pi^\pm\pi^0$ branching ratio, and thus has a small irreducible theoretical 
error. It can reach percent level precision thanks to expected future lattice-QCD 
progress for the calculation of the relevant $SU(3)$-breaking form-factor ratio, and 
serves as a benchmark for new-physics searches. We obtain an interesting picture 
in the $A_{\pi^0 K_{\rm S}}$--$S_{\pi^0 K_{\rm S}}$ plane, where the current 
experimental data show a discrepancy with the Standard Model, and comment on 
the direct CP asymmetries of $B^0\to\pi^-K^+$ and $B^+\to\pi^0K^+$. A modified 
electroweak penguin with a large new CP-violating phase can explain the discrepancy 
and allows us to accommodate also the corresponding data for other $b\to s$ 
penguin-dominated decays.
}
\end{tabular}

\end{center}

\vspace*{1.7truecm}

\vfill

\noindent
June 2008

\end{titlepage}

\newpage
\thispagestyle{empty}
\mbox{}

\newpage
\thispagestyle{empty}
\mbox{}

\rule{0cm}{23cm}

\newpage
\thispagestyle{empty}
\mbox{}

\setcounter{page}{0}

\preprint{CERN-PH-TH/2008-nnn}

\date{June 18, 2008}

\title{\boldmath Benchmarks for the New-Physics Search through
CP Violation in $B^0\to\pi^0 K_{\rm S}$\unboldmath}

\author{Robert Fleischer}
\affiliation{Theory Division, Department of Physics, CERN, CH-1211 Geneva 23,
Switzerland}

\author{Sebastian J\"ager}
\affiliation{Theory Division, Department of Physics, CERN, CH-1211 Geneva 23,
Switzerland}

\author{Dan Pirjol}
\affiliation{National Institute for Physics and Nuclear Engineering, 
Department of Particle Physics, 077125 Bucharest, Romania}

\author{Jure Zupan}
\affiliation{Theory Division, Department of Physics, CERN, CH-1211 Geneva 23,
Switzerland}
\affiliation{J.~Stefan Institute, Jamova 39, 1000 Ljubljana, Slovenia}
\affiliation{Faculty of mathematics and physics, University of Ljubljana, Jadranska 19, 1000 Ljubljana, Slovenia}

\begin{abstract}
\vspace{0.2cm}\noindent
Using isospin relations, we predict the Standard-Model correlation between 
$S_{\pi^0 K_{\rm S}}\equiv (\sin 2\beta)_{\pi^0 K_{\rm S}}$ and $A_{\pi^0 K_{\rm S}}$, the mixing-induced and direct CP asymmetries of $B^0\to \pi^0 K_{\rm S}$. The 
calculation uses flavour $SU(3)$ only to fix the isospin-3/2 amplitude through the 
$B^\pm\to\pi^\pm\pi^0$ branching ratio, and thus has a small irreducible theoretical 
error. It can reach percent level precision thanks to expected future lattice-QCD 
progress for the calculation of the relevant $SU(3)$-breaking form-factor ratio, and 
serves as a benchmark for new-physics searches. We obtain an interesting picture 
in the $A_{\pi^0 K_{\rm S}}$--$S_{\pi^0 K_{\rm S}}$ plane, where the current 
experimental data show a discrepancy with the Standard Model, and comment on 
the direct CP asymmetries of $B^0\to\pi^-K^+$ and $B^+\to\pi^0K^+$. A modified 
electroweak penguin with a large new CP-violating phase can explain the discrepancy 
and allows us to accommodate also the corresponding data for other $b\to s$ 
penguin-dominated decays.
\end{abstract}


\maketitle
Intriguing experimental results for observables of non-leptonic $b\to s$ 
decays \cite{HFAG} have been receiving considerable attention
for several years, where the ``$B\to\pi K$ puzzle'' is an 
important example  (see, e.g., \cite{BFRS-I,BpiK-papers,BeNe,GR,GRZ,FRS}). 
The challenge is to disentangle possible signals of new physics 
(NP) from uncertainties that are related to strong interactions. In this context, a particularly interesting probe is offered by the time-dependent CP asymmetry in 
$B^0 \to \pi^0 K_{\rm S}$, 
\begin{eqnarray}
\lefteqn{\frac{\Gamma(\bar B^0(t)\to  \pi^0 K_{\rm S})\,-\,
\Gamma(B^0(t)\to  \pi^0 K_{\rm S})}{\Gamma(\bar B^0(t)\to  \pi^0 K_{\rm S})\,+
\,\Gamma(B^0(t)\to  \pi^0 K_{\rm S})}}\nonumber\\
&=&A_{\pi^0 K_{\rm S}}\cos(\Delta M_d\,t)+S_{\pi^0 K_{\rm S}}
\sin(\Delta M_d\,t)\,,\label{CPASY}
\end{eqnarray}
where $S_{\pi^0 K_{\rm S}}$ arises from interference between mixing
and decay, and $A_{\pi^0 K_{\rm S}}$ is the ``direct'' CP asymmetry. In the 
Standard Model (SM), we have -- up to doubly Cabibbo-suppressed terms -- the 
following expressions
\cite{PAP-III}:
\begin{equation}
A_{\pi^0 K_{\rm S}}\approx0, \quad
S_{\pi^0 K_{\rm S}}\equiv (\sin 2\beta)_{\pi^0 K_{\rm S}}\approx\sin 2\beta,
\end{equation}
where $\beta$ is one of the angles in the standard unitarity triangle
(UT) of the Cabibbo--Kobayashi--Maskawa (CKM) matrix. 
The current world average is \cite{HFAG}
\begin{equation}\label{s2b-exp-pi0KS}
(\sin 2\beta)_{\pi^0 K_{\rm S}}= 0.58 \pm 0.17,
\end{equation}
which should be compared with the ``reference'' value following from
$B^0\to J/\psi K_{\rm S}$ and similar modes
\begin{equation}\label{s2b-ref}
(\sin 2\beta)_{J/\psi K_{\rm S}}=0.681 \pm 0.025.
\end{equation}

\begin{figure} 
   \centering
   \includegraphics[width=1.5in]{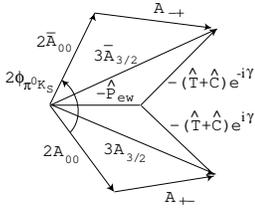} 
   \vspace*{-0.5truecm}
    \caption{The isospin relations \eqref{ampl-rel2} in the complex plane.
    The magnitudes of the amplitudes, $|A_{ij}|\equiv|A(B\to K^i\pi^j)|$ and 
    $|\bar A_{ij}|\equiv|\bar A(B\to K^i\pi^j)|$, can be obtained from the corresponding 
    branching ratios and direct CP asymmetries listed in Table~\ref{tab:data}, while  
    $A_{3/2}$ and $\bar A_{3/2}$ are fixed through \eqref{EWP-SM} and 
    \eqref{T+C-det}.}\label{Fig:1}
 \end{figure}

The search for NP signals in the CP asymmetries of $B^0\to \pi ^0 K_{\rm S}$
requires a reliable SM prediction of $S_{\pi^0 K_{\rm S}}$ and/or $A_{\pi^0 K_{\rm S}}$.
In this letter, we show that $S_{\pi^0 K_{\rm S}}$ can be calculated in the SM
as a function of $A_{\pi^0 K_{\rm S}}$, with projected irreducible theoretical 
errors at the percent level. The starting point  is the isospin relation 
\cite{iso-rel}:
\begin{equation}\label{ampl-rel2}
\begin{array}{c}
\sqrt{2}\,A(B^0\to\pi^0K^0)\,+\,A(B^0\to\pi^-K^+)\\
=-\left[(\hat T+\hat C)e^{i\gamma}\,+\,\hat P_{\rm ew}\right]\equiv 3  A_{3/2};
\end{array}
\end{equation}
a similar relation holds for the CP-conjugate amplitudes, with $A_{3/2}\to \bar A_{3/2}$
and $\gamma \to -\gamma$. Here $\hat T$, $\hat C$ 
and $\hat P_{\rm ew}$ are, respectively, the colour-allowed tree, colour-suppressed 
tree and electroweak penguin (EWP) contributions  \cite{notation}. 
The subscript of $A_{3/2}$ reminds us
that the $\pi K$ final state has isospin $I=3/2$, so that the individual 
QCD penguin contributions cancel in (\ref{ampl-rel2}). 
$S_{\pi^0K_{\rm S}}$ can be written as 
\beq\label{SKSpi0}
S_{\pi^0K_{\rm S}}=\frac{2 |\bar A_{00} A_{00}|}{|\bar A_{00}|^2+|A_{00}|^2}
\sin(2\beta-2\phi_{\pi^0K_{\rm S}}),
\eeq
with $A_{00}\equiv A(B^0\to\pi^0K^0)$ and 
$\bar A_{00}\equiv A(\bar B^0\to\pi^0\bar K^0)$  \cite{BF-98}. 
If $A_{3/2}$ and $\bar A_{3/2}$ are known, 
$2\phi_{\pi^0K_{\rm S}}=\arg(\bar A_{00} A_{00}^*)$ can be fixed through
(\ref{ampl-rel2}), as shown in Fig.~\ref{Fig:1}. In order to determine 
$A_{3/2}$, we first rewrite the 
lower line of (\ref{ampl-rel2}) as
\beq\label{A32}
3  A_{3/2}=-\big(\hat T+ \hat C\big)\big(e^{i\gamma}-q e^{i\omega}).
\eeq
In the SM, the ratio $qe^{i\omega}\equiv -\hat P_{\rm ew}/(\hat T+\hat C)$ is given by
\begin{equation}\label{EWP-SM}
q\,e^{i\omega}=\frac{-3}{2\lambda^2R_b}\frac{C_9(\mu)+
C_{10}(\mu)}{C_1(\mu)+C_2(\mu)}R_q =
0.66\times\frac{0.41}{R_b} R_q,
\end{equation}
where $\lambda\equiv |V_{us}|=0.22$, $R_b=0.41\pm0.04\propto |V_{ub}/V_{cb}|$
is a UT side (value follows from \cite{mannel}), and the $C$s are
Wilson coefficients. If we assume exact $SU(3)$ flavour symmetry and neglect 
penguin contractions, we have $R_q=1$ \cite{BF-98,NR}, while we shall use 
$R_q=1\pm0.3$ for the numerical analysis (results are robust with respect to 
the strong phase $\omega$). Since $qe^{i\omega}$ factorizes at leading order (LO)
in the $1/m_b$ expansion, $R_q$ can be well predicted using factorization techniques 
and future input from lattice QCD.

\begin{table} 
\caption{World averages of experimental data  after ICHEP08 used in the numerical analyses (see also  \cite{HFAG}).}\label{tab:data}
\begin{tabular}{cccc}
\hline\hline
Mode & $\mbox{BR}~[10^{-6}]$ & $A_{\rm CP}$& $ S_{\rm CP}$\\\hline
$\bar B^0\to \pi^+K^-$ & $19.4\pm 0.6$~&~$-0.098\pm0.012$&$-$\\
$\bar B^0\to \pi^0 \bar K^0$ & $9.8 \pm 0.6$ & $-0.01 \pm 0.10$&$0.58\pm0.17$\\
$B^+\to \pi^+\pi^0$ & $5.59\pm0.41$ & $\equiv 0$&$-$\\
$B^0\to \pi^+\pi^-$ & $5.16\pm0.22$ & $0.38\pm 0.06$&$-0.65\pm0.07$\\
$B^0\to \pi^0\pi^0$ & $1.55\pm0.19$ & $0.43\pm 0.25$&$-$\\\hline\hline
   \end{tabular} 
    
 \end{table}

$SU(3)$ flavour symmetry allows us furthermore to fix $|\hat T+\hat C|$ through
the $b\to d$ decay $B^+\to\pi^+\pi^0$ \cite{GRL}:
\begin{equation}\label{T+C-det}
|\hat T+\hat C|=R_{T+C}\left|V_{us}/V_{ud}\right|\sqrt{2}|A(B^+\to\pi^+\pi^0)|,
\end{equation}
where the tiny EWP contributions to $B^+\to \pi^+\pi^0$ were neglected, 
but could be included using isospin \cite{BF-98,GPY}. We stress 
that (\ref{T+C-det}) does {\it not} rely on further dynamical assumptions.
For the $SU(3)$-breaking parameter $R_{T+C}\sim f_K/f_\pi$ we use 
the value $1.22\pm0.2$, where the error is quite conservative, as discussed below.

\begin{figure}
    \centering
    \begin{tabular}{cc}
   \includegraphics[width=1.5in]{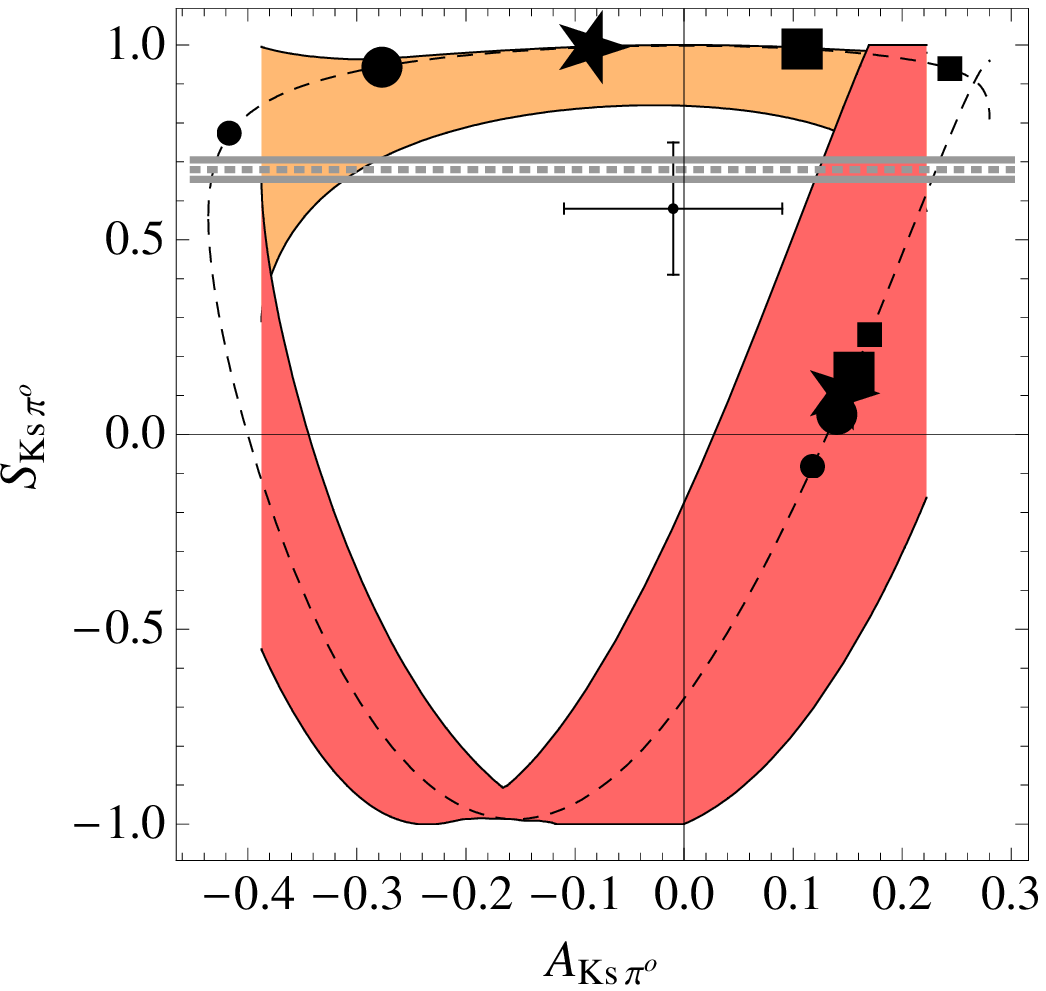} &
   \includegraphics[width=1.5in]{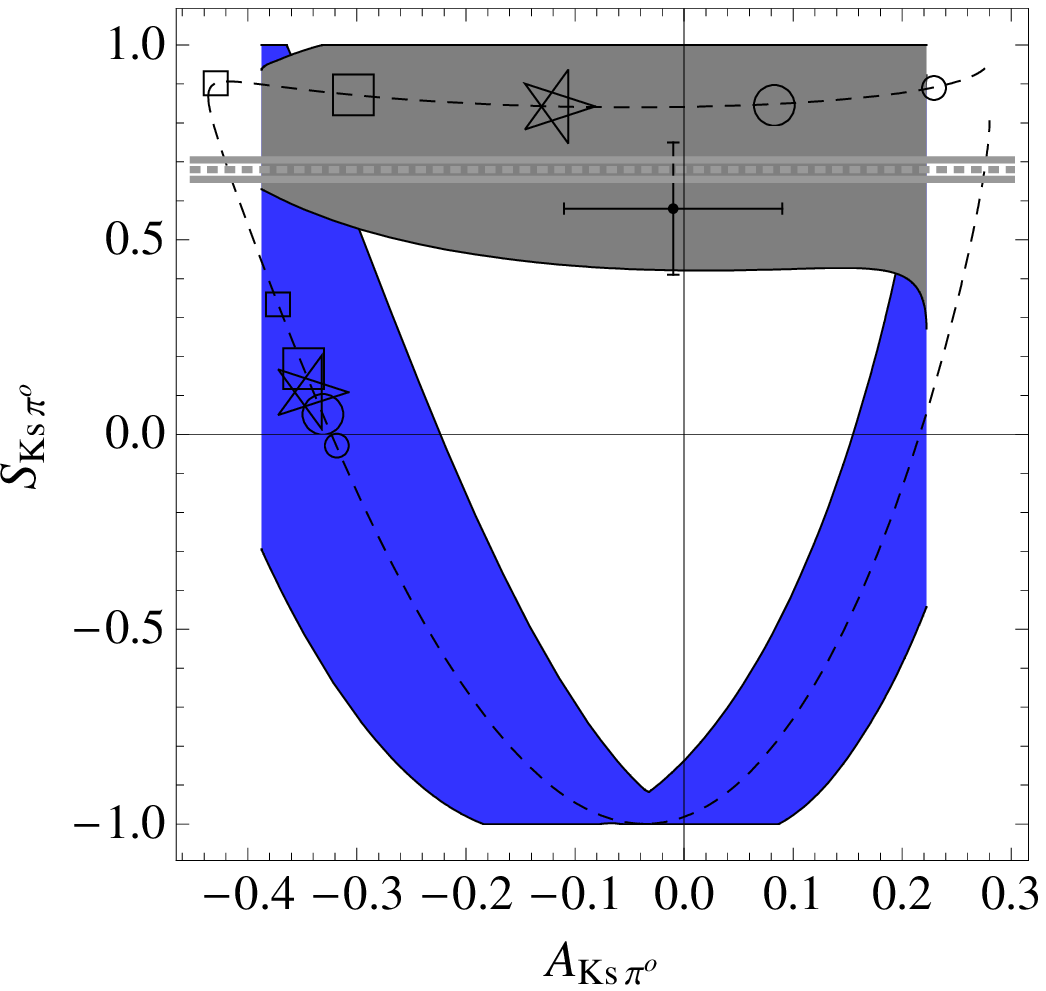}
   \end{tabular} 
    \caption{The SM constraints in the $A_{\pi^0K_{\rm S}}$--$S_{\pi^0K_{\rm S}}$ plane,
    as explained in the text. Left panel: contains $\delta_{\rm c} \approx 0^\circ$ 
    (consistent with QCD), with
    $\delta_{\rm c}=-60^\circ$ (small circle), $-30^\circ$ (large circle),
$0^\circ$ (star), $30^\circ$ (large square), $60^\circ$ (small square).
    Right panel: contains $\delta_{\rm c}\approx 180^\circ$ (not consistent with QCD),
    with $\delta_{\rm c}=120^\circ$ (small circle), $150^\circ$ 
(large circle), $180^\circ$ (star), $210^\circ$ (large square), 
$240^\circ$ (small square). The shaded horizontal bands represent 
the value of $(\sin2\beta)_{J/\psi K_{\rm S}}$ in (\ref{s2b-ref}).}\label{Fig:2}
 \end{figure}

Relations \eqref{A32}--\eqref{T+C-det} allow us to determine $A_{3/2}$ and 
$\bar A_{3/2}$, thereby fixing  the two isospin triangles in Fig.~\ref{Fig:1}. 
Since the triangles can be flipped around the $A_{3/2}$ and $\bar A_{3/2}$ sides, 
we encounter a fourfold ambiguity (not shown). Using  \eqref{SKSpi0}, 
$S_{\pi^0K_{\rm S}}$ is determined as well. The corresponding prediction 
is shown in Fig.~\ref{Fig:2}, where we keep $A_{\pi^0K_{\rm S}}$ as a free parameter. 
For the implementation of this construction, we express the curves 
in Fig.~\ref{Fig:2} in parametric form \cite{BFRS-I} as functions of a 
strong phase $\delta_c$, defined through
\beq
r_{\rm c}e^{i\delta_{\rm c}}=\big(\hat T+\hat C\big)/\hat P,
\eeq
where $\hat P$ is the $B^0 \to \pi^- K^+$ penguin amplitude \cite{notation}. 
We find that no solutions exist for certain ranges of $\delta_{\rm c}$,
separating the full $[0^\circ,360^\circ]$ range into two regions. They contain 
$\delta_{\rm c}=0^\circ$ or $180^\circ$ and correspond to the left and
right panels of Fig.~\ref{Fig:2}, respectively. As one circles the trajectory in either 
panel by changing  $\delta_{\rm c}$, each value of this strong phase in the respective 
interval is attained twice. In order to illustrate this feature, we show -- for 
central values of the input data/parameters -- points corresponding
to various choices of $\delta_{\rm c}$. 
The bands show the $1\,\sigma$ variations obtained by 
adding  in quadrature the errors due to all input data/parameters.
Moreover, we assume $\gamma=65^\circ\pm10^\circ$ \cite{UTfit,CKMfitter}. 
This angle will be determined with 
excellent accuracy thanks to CP violation measurements in pure tree $B$ decays 
at the LHCb experiment (CERN).

\begin{figure} 
   \centering
       \begin{tabular}{cc}
     \includegraphics[width=1.5in]{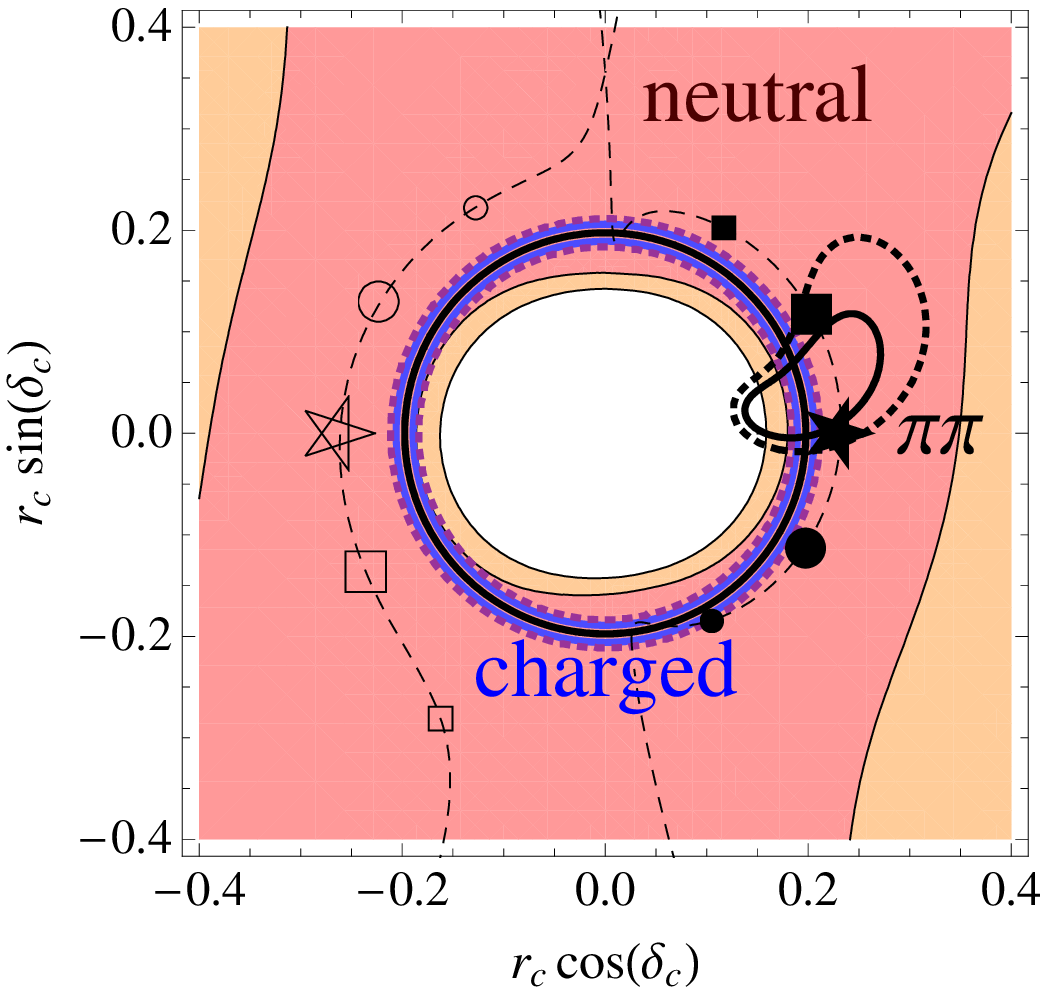}
       \includegraphics[width=1.5in]{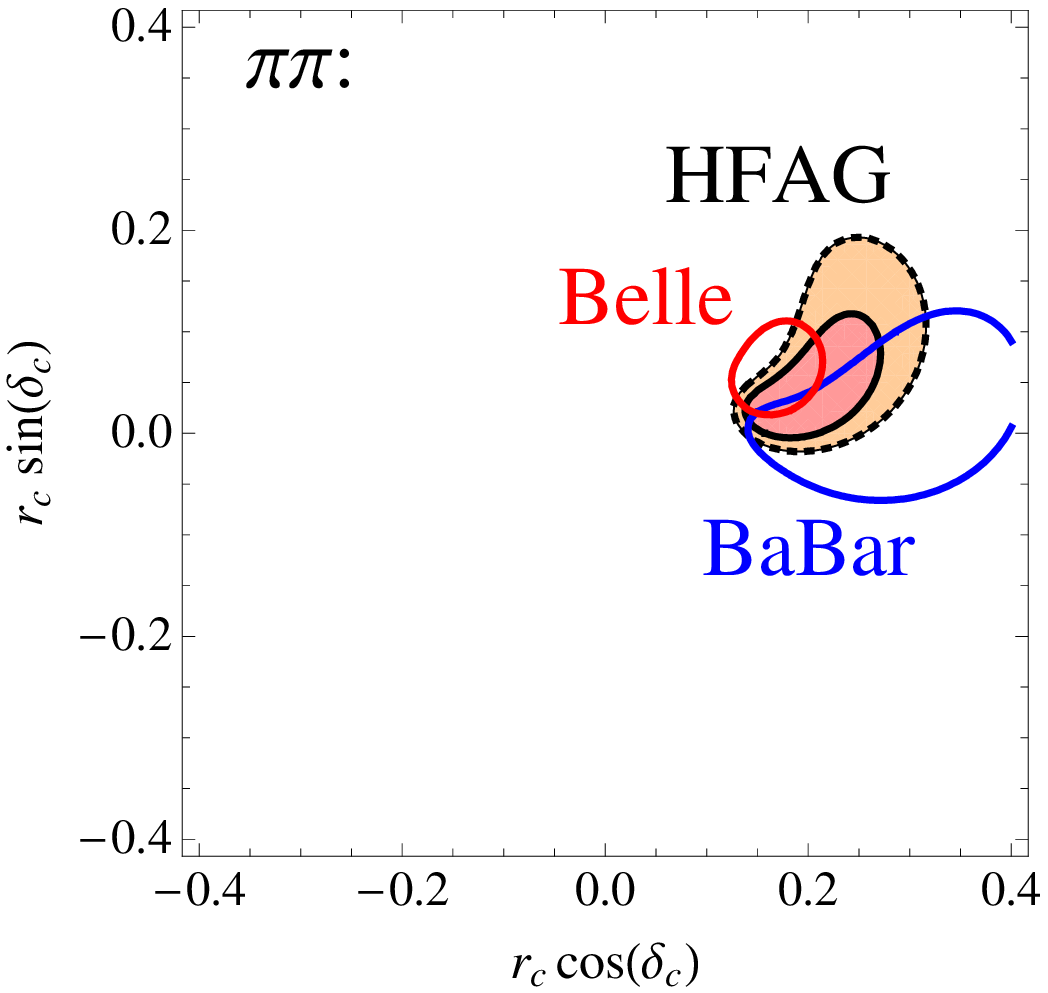}
   \end{tabular} 
   \caption{The constraints on $r_{\rm c}e^{i\delta_{\rm c}}$ that 
   follow from the current data, as discussed
   in the text. Left panel:  $B\to\pi K$ and $B\to\pi\pi$ constraints (the symbols 
   to label $\delta_{\rm c}$ correspond to those in Fig.~\ref{Fig:2}). Right panel: 
   $B\to\pi\pi$ constraints for the BaBar and Belle data 
   for $A_{\pi^+\pi^-}$ and the HFAG average. The solid and dotted lines 
   refer to $1\,\sigma$ and 90\% C.L. ranges,
   respectively.}\label{Fig:3}
\end{figure}

In order to resolve the fourfold ambiguity in Fig.~\ref{Fig:2}, we need further 
information on $r_{\rm c}$, $\delta_{\rm c}$: i) $r_{\rm c}$ can be determined 
if we fix $|\hat T+\hat C|$ through $\mbox{BR}(B^+\to \pi^+\pi^0)$ (see \eqref{T+C-det}) 
and $|\hat P|$ through $\mbox{BR}(B^+\to\pi^+K^0)\propto |\hat P|^2+\dots$, where 
the dots represent negligible doubly Cabibbo-suppressed terms that are already
strongly constrained by data \cite{BKK}. In the left panel of Fig.~\ref{Fig:3}, 
the corresponding $r_{\rm c}$ constraint is shown at the ``charged'' circle. ii) 
Using the $SU(3)$ flavour symmetry and other plausible dynamical assumptions
\cite{BFRS-I}, a fit to all available $B\to\pi\pi$ data yields the $\pi\pi$ curves. 
Since BaBar and Belle do not fully agree on the measurement of the direct 
CP asymmetry in $B^0\to\pi^+\pi^-$ \cite{HFAG}, we show in the right panel of  
Fig.~\ref{Fig:3} the corresponding allowed regions separately. We observe that 
the data imply $\delta_{\rm c}\sim(0\mbox{--}30)^\circ$, in agreement with the 
heavy-quark expansion analyses in \cite{BeNe,BBNS} and \cite{SCET}, differing 
in their treatment of non-perturbative charm-penguin contributions. Consequently, 
we can exclude the solutions shown in the right panel of Fig.~\ref{Fig:2}, and are left 
with the twofold solution in the left panel. However, the lower band corresponds
to $r_{\rm c}$ values of the ``neutral'' region in the left panel of Fig.~\ref{Fig:3}
that are far off the right of the displayed region, drastically inconsistent both 
with the $B\to\pi\pi$ data and with the heavy-quark limit.

\begin{figure}
   \centering
   \includegraphics[width=1.5in]{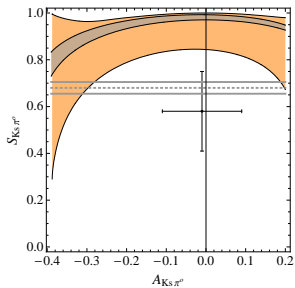} 
   \caption{The correlation in the $A_{\pi^0K_{\rm S}}$--$S_{\pi^0K_{\rm S}}$ 
   plane for a future benchmark scenario (narrow band) in comparison with the 
   current situation (wider band), as explained in the text.}
   \label{Fig:4}
\end{figure}

 Consequently, we are left with the thin horizontal part of the upper band in 
 the left panel of Fig.~\ref{Fig:2}, which we show enlarged in Fig.~\ref{Fig:4}. 
 Using the experimental value for $A_{\pi^0K_{\rm S}}$, we obtain the SM 
 prediction 
 \beq\label{SM-pred}
S_{\pi^0K_{\rm S}}=0.99^{+0.01}_{-0.08}\bigl|_{\rm exp.}{}^{+0.000}_{-0.001}
\big|_{R_{\rm T+C}}{}^{+0.00}_{-0.11}\bigl|_{R_{q}}{}^{+0.00}_{-0.07}
\bigl|_{\gamma}, 
\eeq
which is about two standard deviations away from the experimental
result in (\ref{s2b-exp-pi0KS}). It should be noted that (\ref{SM-pred}) 
depends on the input data collected in Table~\ref{tab:data}.

In Fig.~\ref{Fig:4}, we show the future theory error benchmark for the
SM constraint in the $A_{\pi^0K_{\rm S}}$--$S_{\pi^0K_{\rm S}}$ plane.
Both $R_q$ \eqref{EWP-SM} and $R_{T+C}$ \eqref{T+C-det} factorize at  LO
in the $1/m_b$ expansion, and can be well predicted using input from lattice 
QCD. It should be stressed that ``charming penguins'' do not enter these ratios. 
As a working tool we use the approach of Ref.~\cite{BeNe,BBNS} (BBNS), but 
similar conclusions can be reached using Ref.~\cite{SCET} (where also 
derivatives of form factors would be needed). The
key parameter is $R_q$, which dominates the current theoretical error  
\eqref{SM-pred}. Its uncertainty is governed by the $SU(3)$-breaking 
form-factor ratio $\xi_{\pi K}\equiv F^{B\to K}(0)/F^{B\to\pi}(0)$. If we assume $\xi_{\pi K}=1.2(1\pm0.03)$, i.e.\ a $20\%$ determination of the 
$SU(3)$-breaking corrections, as an optimistic -- but achievable -- goal for lattice 
QCD, we obtain the BBNS result 
$R_q = (0.908^{+0.052}_{ - 0.043}) e^{i (0^{+ 1}_{ - 1})^\circ}$, to be compared with 
the present value $R_q=(1.02^{+0.27}_{-0.22}) e^{i (0^{+ 1}_{ - 1})^\circ}$ 
\cite{Beneke:2006mk}.
Similarly, we find $R_{T+C}=1.23^{+0.02}_{-0.03}$, where the increase of
precision is very mild as the form-factor dependence essentially cancels out.
Setting, moreover, the uncertainties of the experimental inputs to zero, 
while keeping central values fixed, we obtain a prediction of $S_{\pi^0K_{\rm S}}$ 
with errors at the percent level, as shown in Fig.~\ref{Fig:4}. Consequently, the 
irreducible theory error of our proposed method for predicting $S_{\pi^0K_{\rm S}}$ 
in the SM is much smaller than in calculations using only the $1/m_b$ expansion,
and makes it promising for a future $e^+e^-$ super-$B$ factory (for a review,
see, e.g., Ref.~\cite{Browder:2008em}). 

\begin{figure}
   \centering
   \includegraphics[width=1.5in]{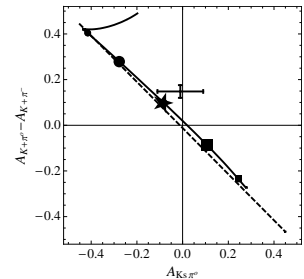} 
   \caption{The SM correlation between $A_{\pi^0K^+}-A_{\pi^-K^+}$ and
   $A_{\pi^0K_{\rm S}}$ for central values of inputs, with hadronic parameters fixed as for Fig. \ref{Fig:2} (solid), or following from the sum rule for rate differences \cite{Gronau:2006xu} (dashed). The dependence on $\delta_c$  is as in Fig. \ref{Fig:2} and is constrained to SM values (upper curve in Fig. \ref{Fig:2}a).
}
   \label{Fig:5}
\end{figure}

Before turning to the interpretation of the current experimental data in terms of NP, let us
briefly comment on the difference of direct CP asymmetries $A_{\pi^0K^+}-A_{\pi^-K^+}$, 
which recently received
quite some attention as a possible sign of NP \cite{belle-nature}.
Fig.~\ref{Fig:5} shows the SM correlation between this difference and the
CP asymmetry $A_{\pi^0K_{\rm S}}$, keeping $A_{\pi^-K^+}$
fixed. 
It depends on CP-averaged $B\to\pi K$ branching ratios
and $\gamma$, and becomes equivalent to the sum rule for rate
differences \cite{Gronau:2006xu} when neglecting higher orders in
subleading amplitudes.
We see that
current data
(cross) can be accommodated 
in the SM within the error on $A_{\pi^0K_{\rm S}}$, although
hadronic amplitudes then deviate from the $1/m_b$ pattern
(see also Ref.~\cite{FRS}).
It would be
desirable to reduce this uncertainty
in the future.

\begin{figure} 
    \centering
    \begin{tabular}{cc}
   \includegraphics[width=1.5in]{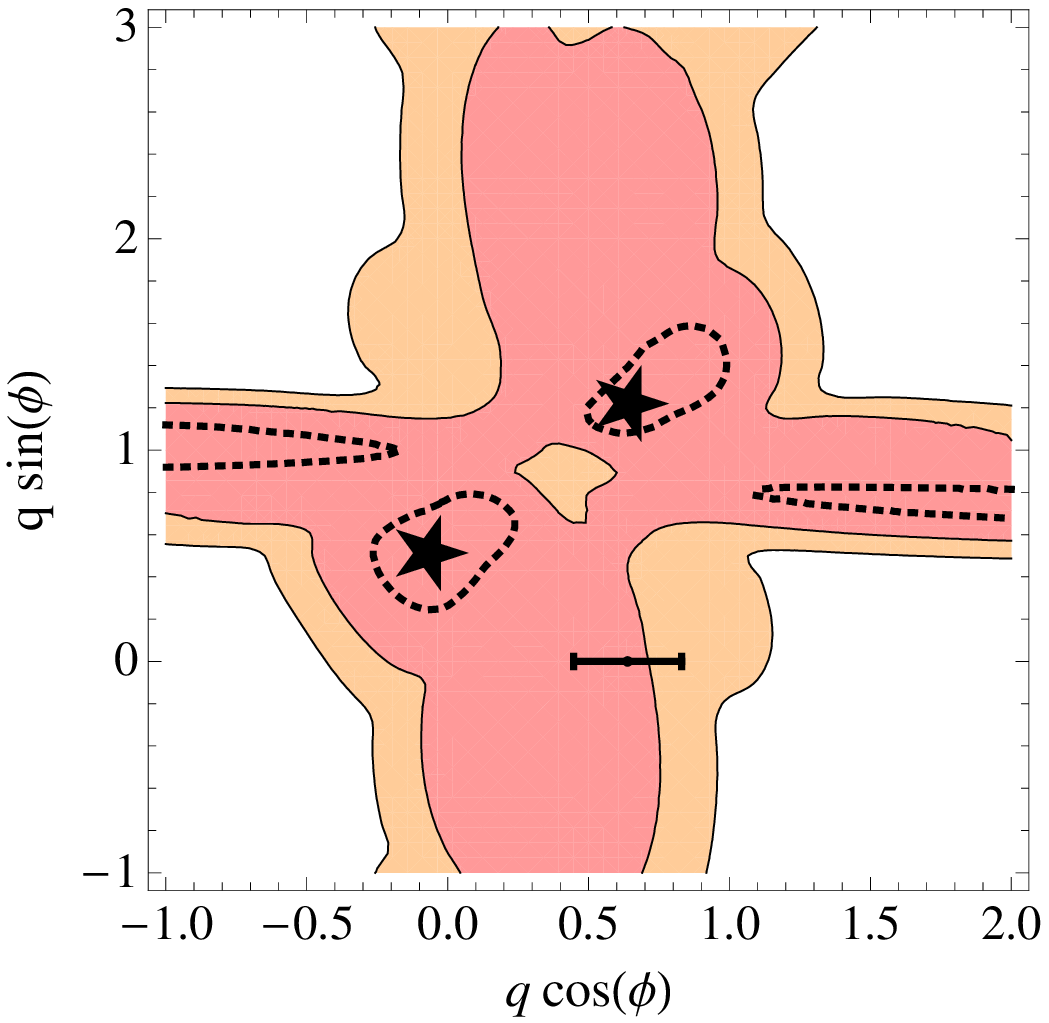} &
   \includegraphics[width=1.5in]{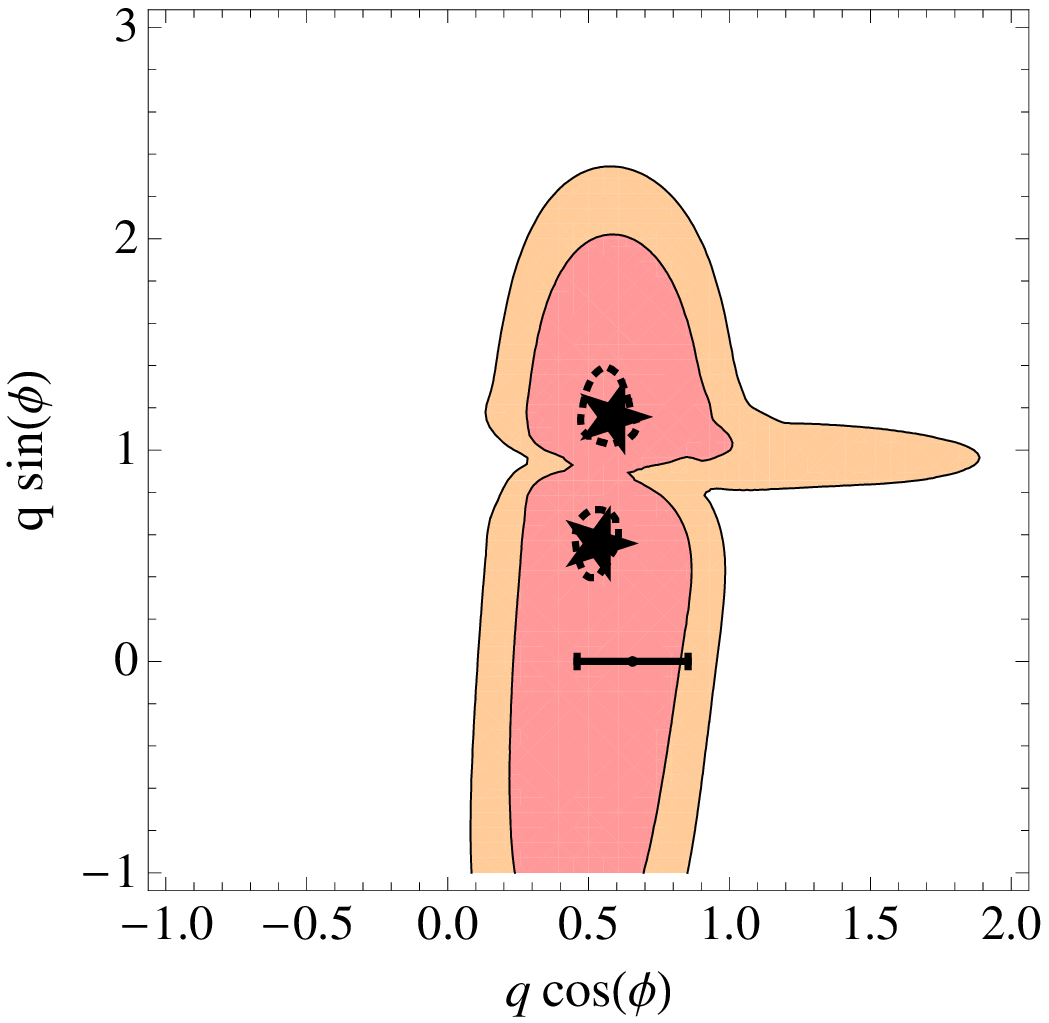}
   \end{tabular} 
    \caption{Constraints on $qe^{i\phi}$. Left panel: $\chi^2$ fit, using only
    the $B\to\pi K$ data. Right panel: $\chi^2$ fit, using both the $B\to\pi K$ and
    $B\to\pi\pi$ data. The inner and outer regions correspond to $1\,\sigma$ and 
    $90\%$ C.L., respectively, while the stars denote the minima of the fits.  
    The  90\% C.L. regions with 10 times more data lie inside the dotted lines 
    (see also the text).}\label{Fig:6}
 \end{figure}

Let us now consider a NP scenario, which allows us to resolve the discrepancy
between (\ref{s2b-exp-pi0KS}) and (\ref{SM-pred}). Following \cite{BFRS-I}, 
we assume that NP manifests itself effectively in the data as a modified EWP 
with a CP-violating NP phase $\phi$, i.e.
$q\to q e^{i\phi}$ in \eqref{A32}. Here $q$ can differ from the SM value in 
(\ref{EWP-SM}). Since $\delta_{\rm c}$ is rather small, the impact of this type of NP on 
$A_{\pi^0K_{\rm S}}$ and $A_{\pi^0K^+}$ is suppressed. In Fig.~\ref{Fig:6}, we show constraints on $qe^{i\phi}$ from two $\chi^2$ fits, using only the $B\to\pi K$ data or 
both the $B\to\pi K$ and $B\to\pi\pi$ data.
The latter have a strong impact on the allowed region 
of $q e^{i\phi}$ \cite{BFRS-I,FRS}, yielding two almost degenerate minima,
$q=1.3\pm0.4$, $\phi=(63^{+10}_{-9})^\circ$ and $q=0.8^{+0.2}_{-0.3}$, $\phi=(45^{+18}_{-28})^\circ$. We also show 
the $90\%$ C.L. regions (dashed curves) that correspond to a future scenario,
assuming the benchmark value of $R_q$ used in Fig.~\ref{Fig:4} and 
ten-times more data, with central values fixed to the present 
$\chi^2$ minimum. In the $\chi^2$ fits we allow all ratios of $SU(3)$-related 
amplitudes to fluctuate flatly around $f_K/f_\pi$ within $30 \%$ in magnitude 
and $30^\circ$ in phase.

\begin{figure} 
    \centering
    \begin{tabular}{cc}
    \includegraphics[width=1.5in]{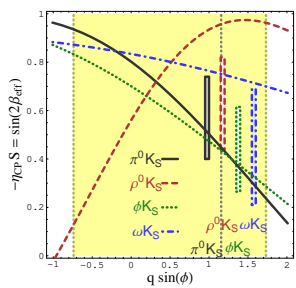}
   \end{tabular} 
    \caption{Mixing-induced CP asymmetries for a set of
penguin-dominated $B^0$ decays as functions of $q \sin(\phi)$,
with $q \cos(\phi)$ fixed to $0.6$. The vertical bars depict the experimental 
$1\,\sigma$ ranges \cite{HFAG}. The $1\,\sigma$ range (vertical band) and best-fit values (dashed line) for $q \sin \phi$ from Fig.~\ref{Fig:6} are also shown.
\label{Fig:7}}
\end{figure}

The possibility of resolving the discrepancy between (\ref{s2b-exp-pi0KS}) and 
(\ref{SM-pred}) through a modified EWP is intriguing. We next illustrate that the 
observed pattern of the mixing-induced 
CP asymmetries in other penguin-dominated $b\to s$ decays \cite{HFAG} can also be
accommodated in the same NP scenario. In Fig.~\ref{Fig:7}, we show the 
results of a BBNS calculation of the $S$ parameters  for four channels of this kind:
we assume that all electroweak Wilson coefficients are rescaled by the same factor 
$qe^{i \phi}$, and use as input the preferred data set ``G'' of \cite{Beneke:2006mk}. 
The value of $qe^{i\phi}$ is then varied along a contour that runs vertically through the preferred region in Fig.~\ref{Fig:6}. Unlike the SM,  the modified EWP scenario
allows us to accommodate the data well (see, e.g., also \cite{FRS,sparSM}). 
The same is true for a more specific scenario where the effective FCNC couplings
of the $Z$ boson at the weak scale are suitably modified. 
Since $S_{ \eta'K_{\rm S}}$ receives a tiny, negative shift from $\sin 2\beta$, 
in agreement with the data, we do not show this in Fig.~\ref{Fig:7}.

In conclusion, we have demonstrated that the SM correlation in the
$A_{\pi^0K_{\rm S}}$--$S_{\pi^0K_{\rm S}}$ plane can be predicted
reliably in the SM, with small irreducible theoretical errors, and have
shown that the resolution of the present discrepancy with the data 
can be achieved through a modified EWP sector, with a large CP-violating
NP phase.

\noindent{\it Acknowledgements} \\
We would like to thank D. Becirevic, M. Della Morte and A. Kronfeld for 
useful discussions of lattice QCD. S.J. is supported in part by the RTN European Program MRTN- 
CT-2004-503369.


\begin{thebibliography}{99}
%
%
%
\bibitem{HFAG} E. Barbiero {\it et al.} [Heavy Flavour Averaging Group
Collaboration], arXiv:0704.3575; updates are available at
{\tt http://www.slac.stanford.edu/xorg/hfag/}.


  \bibitem{BFRS-I}A.~J.~Buras {\it et al.}
  {\it Phys.\ Rev.\ Lett.}~{\bf 92},  101804  (2004);
  {\it Nucl.\ Phys.}~{\bf B697}, 133 (2004).


\bibitem{BpiK-papers}T.~Yoshikawa,
  {\it Phys.\ Rev.}~{\bf D68},  054023 (2003);
  M.~Gronau and J.~L.~Rosner,
  {\it Phys.\ Lett.}~{\bf B572}, 43 (2003); 
  A.~J.~Buras {\it et al.},
  {\it Eur.\ Phys.\ J.}~{\bf C32}, 45 (2003);
  V.~Barger {\it et al.},
  {\it Phys.\ Lett.}~{\bf B598}, 218 (2004);
   Y.~L.~Wu and Y.~F.~Zhou,
  {\it Phys.\ Rev.}~{\bf D72}, 034037 (2005);
  T.~Feldmann {\it et al.},
  arXiv:0803.3729 [hep-ph],
  S.~Baek {\it et al.},
  Phys.\ Rev.\  D {\bf 71}, 057502 (2005)
  S.~Baek and D.~London,
  Phys.\ Lett.\  B {\bf 653}, 249 (2007),
  K.~Agashe {\it et al.},
  arXiv:hep-ph/0509117.




  
 \bibitem{BeNe}M.~Beneke and M.~Neubert,
  {\it Nucl.\ Phys.}~{\bf B675}, 333 (2003).

\bibitem{GR}M.~Gronau and J.~L.~Rosner,
  {\it Phys.\ Rev.}~{\bf D71}, 074019 (2005);
  {\it Phys.\ Lett.}~{\bf B644}, 237 (2007).
  
  \bibitem{GRZ}M.~Gronau {\it et al.}
  {\it Phys.\ Rev.}~{\bf D74}, 093003 (2006).
  
\bibitem{FRS}R.~Fleischer {\it et al.},
 {\it Eur.\ Phys.\ J.}~{\bf C51}, 55 (2007).

\bibitem{PAP-III}R.~Fleischer,
  {\it Phys.\ Lett.}~{\bf B365}, 399 (1996).

\bibitem{iso-rel}Y.~Nir and H.~R.~Quinn, {\it Phys.\ Rev.\ Lett.}~{\bf 67},
541 (1991); M.~Gronau {\it et al.},
{\it Phys.\ Rev.}~{\bf D52}, 6374 (1995).
 

\bibitem{notation}
We are using a notation very similar to \cite{BFRS-I}, with 
$\hat T\equiv |V_{ub}V_{us}|{\cal T}'$, $\hat C\equiv |V_{ub}V_{us}|{\cal T}'$
and $\hat P\equiv |V_{cb}V_{cs}|({\cal P}_t'-{\cal P}_c')$, while the quantities
$q$, $\omega$, $r_{\rm c}$ and $\delta_{\rm c}$ agree with \cite{BFRS-I}.


\bibitem{BF-98}A.J. Buras and R.~Fleischer,
  {\it Eur.\ Phys.\ J.}~{\bf C11}, 93 (1999).

\bibitem{NR}M.~Neubert and J.~L.~Rosner,
  {\it Phys.\ Rev.\ Lett.}~{\bf 81}, 5076 (1998).

\bibitem{mannel}T. Mannel, seminar given at CERN, May 8th, 2008.

\bibitem{GRL}M.~Gronau {\it et al.}
{\it Phys.\ Rev.\ Lett.}~{\bf 73}, 21 (1994).


\bibitem{GPY}M.~Gronau {\it et al.},
  {\it Phys.\ Rev.}~{\bf D60} (1999) 034021
  [Erratum-ibid.\  {\bf D69} (2004) 119901].

\bibitem{UTfit}M.~Bona {\it et al.}~[UTfit Collaboration],
  {\it JHEP} {\bf 0507}, 028 (2005); updates: 
{\tt http://utfit.roma1.infn.it/}.

\bibitem{CKMfitter}J.~Charles {\it et al.}~[CKMfitter Group], 
{\it Eur.\ Phys.\ J.}~{\bf C41}, 1 (2005); updates: {\tt http://ckmfitter.in2p3.fr/}.

\bibitem{BKK}R.~Fleischer,
  {\it Eur.\ Phys.\ J.}~{\bf C52}, 267 (2007).
  
 \bibitem{BBNS}M.~Beneke {\it et al.},
  {\it Phys.\ Rev.\ Lett.}~{\bf 83}, 1914 (1999).


\bibitem{SCET}
  C.~W.~Bauer {\it et al.},
 {\it Phys.\ Rev.}~{\bf D70}, 054015 (2004),
  {\it Phys.\ Rev.}~{\bf D74}, 034010 (2006);
  A.~R.~Williamson and J.~Zupan,
  {\it Phys.\ Rev.}~{\bf D74}, 014003 (2006).


\bibitem{Beneke:2006mk}
  M.~Beneke and S.~J\"ager,
 {\it Nucl.\ Phys.}~{\bf B768}, 51 (2007).
  
\bibitem{Browder:2008em}
  T.~E.~Browder {\it et al.},
  arXiv:0802.3201 [hep-ph].

\bibitem{belle-nature} S.-W. Lin {\it et al.}  [Belle Collaboration], 
{\it Nature} {\bf 452}, 332 (2008).




\bibitem{Gronau:2006xu}
  M.~Gronau and J.~L.~Rosner,
 {\it Phys.\ Rev.}~{\bf D74}, 057503 (2006).

\bibitem{sparSM}
  G.~Buchalla {\it et al.},
  {\it JHEP} {\bf 0509}, 074 (2005).
  M.~Beneke,
  {\it Phys.\ Lett.}~{\bf B620}, 143 (2005).

%
%
%
\end{thebibliography}
\end{document}